\documentclass[aps,prl,twocolumn,superscriptaddress,showpacs,floatfix]{revtex4}

\usepackage{graphicx}

\begin{document}

\title{Singlet-triplet dispersion reveals additional frustration in the triangular dimer compound Ba$_3$Mn$_2$O$_8$}

\author{M. B. Stone}
\affiliation{Neutron Scattering Science Division, Oak Ridge National
Laboratory, Oak Ridge, Tennessee 37831, USA}

\author{M. D. Lumsden}
\affiliation{Neutron Scattering Science Division, Oak Ridge National
Laboratory, Oak Ridge, Tennessee 37831, USA}

\author{S. Chang}
\affiliation{NIST Center for Neutron Research, Gaithersburg,
Maryland 20899, USA}

\author{E. C. Samulon}
\affiliation{Department of Applied Physics and Geballe Laboratory
for Advanced Materials, Stanford University, California
94305, USA}

\author{C. D. Batista}
\affiliation{Theoretical Division, Los Alamos National Laboratory, Los Alamos, New Mexico, 87545 USA}

\author{I. R. Fisher}
\affiliation{Department of Applied Physics and Geballe Laboratory
for Advanced Materials, Stanford University, California
94305, USA}

\begin{abstract}
We present single crystal inelastic neutron scattering measurements of
the $S=1$ dimerized quasi-two-dimensional antiferromagnet Ba$_3$Mn$_2$O$_8$.
The singlet-triplet dispersion reveals nearest-neighbor and next-nearest-neighbor ferromagnetic interactions between adjacent bilayers that compete against each other. Although the inter-bilayer exchange is comparable to the intra-bilayer exchange, this additional frustration reduces the effective coupling along the $c$-axis and leads to a quasi-two dimensional behavior. In addition, the obtained exchange values are able to reproduce the four critical fields in the phase diagram.
\end{abstract}


\maketitle
The presence of exotic ground states \cite{gs}, excitations \cite{excitations}
and quantum phase transitions \cite{qpt}
in a growing  number of available materials is making low-dimensional gapped frustrated quantum
magnets especially topical. In some cases, the ability to tune the system Hamiltonian via external fields
allows access to additional phases. For instance, the
singlet-triplet gap in the spectrum can be closed via application of external
magnetic fields \cite{rueggnature2003,stonenjp}. The resulting quantum critical point (QCP)
separates a low field disordered paramagnet from a high field ordered
phase which to a good approximation can be considered a Bose-Einstein condensation of
magnons \cite{affleck1990,nikuniprl}.

\begin{figure}[tb!]
\centering\includegraphics[scale=0.4]{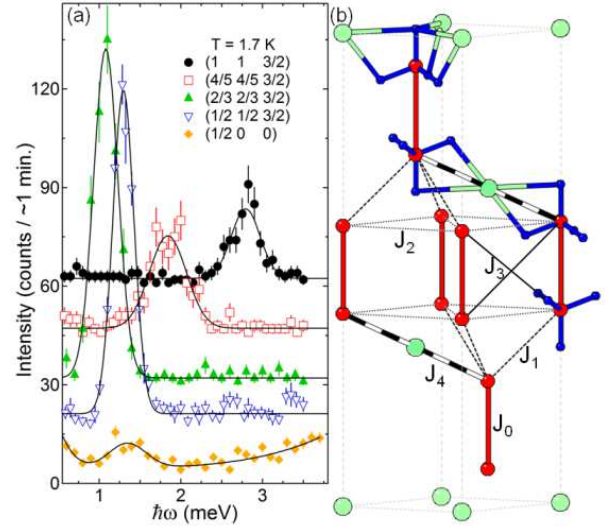}
\caption{\label{fig:fig1}(a) Constant $\mathbf{Q}$ scans
in the $(hhl)$ and $(hk0)$ planes.  Solid lines are
Gaussian fits.  The $(\frac{1}{2}00)$ fit includes a
fixed elastic Gaussian and a higher energy Gaussian to account
for incoherent elastic scattering and a scattering angle dependent background respectively.  Data are offset vertically
for presentation. (b) Crystal structure of Ba$_3$Mn$_2$O$_8$ showing
Mn$^{5+}$ (red) and 12-fold coordinated Ba sites (green)\cite{weller1999}.
Oxygen sites (blue) involved in coordination of two dimers are shown to illustrate
interdimer connectivity and tetrahedral coordination of the Mn$^{5+}$.  10-fold coordinated Ba sites are not shown for clarity.  Exchange connectivity
within and between $S=1$ dimers is shown with different line types.  $J_3$ is only shown for a single pair of dimers.  Exchange constants correspond to Mn-Mn
distances of 3.985~\AA ($J_0$), 4.569~\AA ($J_1$), 5.711~\AA ($J_2$), 6.964~\AA ($J_3$) and 7.313~\AA ($J_4$).}
\end{figure}

The hexagonal antiferromagnetic bilayer Ba$_3$Mn$_2$O$_8$ is established as a quasi two
dimensional (2D) frustrated antiferromagnet with a quantum critical phase diagram\cite{uchidajphys2001,samulonbamno}.
Rather than a single Bose-Einstein condensate phase as in $S=\frac{1}{2}$ singlet-triplet antiferromagnets,
Ba$_3$Mn$_2$O$_8$ has two sequential magnetically ordered phases, \textit{i.e.} four zero-temperature QCPs,
as a function of applied magnetic field: $\mu_0 H_{c1} \approx 9$~T,
$\mu_0 H_{c2} \approx 26$~T, $\mu_0 H_{c3} \approx 32.3$~T and
$\mu_0 H_{c4} \approx 48$~T\cite{uchidaprb2002,Tsujiiprb2005,xuprb2008}.
The first (second) ordered phase for $H_{c1} < H < H_{c2}$ ($H_{c3} < H < H_{c4}$) can be approximately described as singlet-triplet
(triplet-quintet) condensation.  The nature of these high field ordered phases fundamentally depends upon the zero field exchange paths and constants.

Powder INS measurements of Ba$_3$Mn$_2$O$_8$ yielded three independent exchange constants through comparison to a spherically averaged scattering intensity\cite{stonebamno}.  The dimer exchange
$J_0$ and interdimer exchanges are illustrated in Fig.~\ref{fig:fig1}(b).  Examining these excitations
using single crystal INS, we find next nearest neighbor (NNN) interactions $J_4$ play a significant and
heretofore unexplored role in the propagation of triplet excitations in Ba$_3$Mn$_2$O$_8$.  Since the
dispersion bandwidth only depends on the difference $J_2-J_3$, we use $\mu_0 H_{c1}$ \cite{uchidaprb2002,Tsujiiprb2005,xuprb2008}
to determine the $J_3$ exchange.  We find that both inter-bilayer
exchange constants, $J_1$ and $J_4$, are important in determining the critical fields. In addition, we show
that although ferromagnetic, $J_1$ and $J_4$ are competing interactions due to the geometric frustration of the triangular bilayers and the inter-bilayer exchange paths. This explains why Ba$_3$Mn$_2$O$_8$ is a quasi-2D magnet in spite of the fact that the dominant inter-bilayer coupling, $J_1$, is larger than $|J_2|$.

\begin{figure*}[htb!]
\centering\includegraphics[scale=0.875]{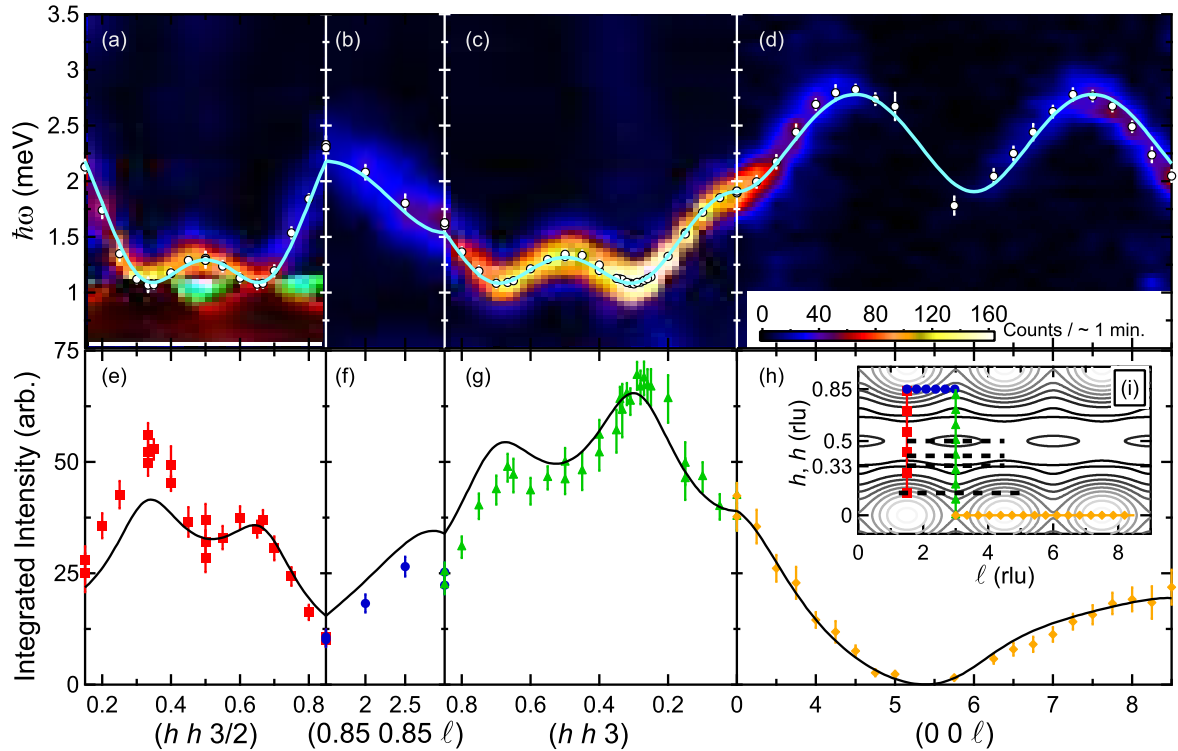}
\caption{\label{fig:fig2}(a)-(d) $T=1.7$~K scattering intensity of Ba$_3$Mn$_2$O$_8$ versus $\hbar\omega$ and $\mathbf{Q}$.  Figure is interpolated from constant $\mathbf{Q}$ scans
in the $(hhl)$ plane.  White circles are peak positions from Gaussian fits described in the text.
Solid line is the fitted dispersion.  (e)-(h) Integrated scattering intensity from Gaussian fits to
constant $\mathbf{Q}$ scans.  Solid line is the fitted scattering intensity.  (i) Path through the $(hhl)$
plane shown in (a)-(h).  Dashed lines correspond to scans shown in Fig.~\ref{fig:fig3}.
Contour lines are the fitted dispersion with dark[light] lines representing smaller[larger] energy transfers shown at 0.2 meV intervals.  Horizontal axes of (a)-(h) are plotted in reciprocal lattice units (rlu).}
\end{figure*}

Single crystals of Ba$_3$Mn$_2$O$_8$ were grown from a flux melt as described in Ref.~\onlinecite{samulonbamno}.
Five crystals with a net mass 0.372 g were coaligned using the HB1A triple-axis spectrometer (TAS) at the High Flux Isotope Reactor with a resulting mosaic of $\approx 42^{\prime}$.
INS measurements were performed on the SPINS cold neutron TAS at the NIST Center
for Neutron Research.  The instrument was configured with $80^{\prime}$ collimation before the sample followed by a cooled Be filter and an $80^{\prime}$ radial collimator.  Measurements were performed with fixed final energy $E_f=5$~meV selected by a $10.5$~cm wide horizontally focussing analyzer.  This configuration results in a measured elastic full width at half maximum (FWHM) energy resolution at the
$(0.3~0.3~3)$ wave vector of $\delta \hbar \omega =0.294(7)$~meV.  The majority of our data presented here consists of constant wave vector $\mathbf{Q}$ scans performed at $T=1.7$~K in the $(hhl)$ scattering plane.

Representative constant $\mathbf{Q}$ scans are shown in Fig.~\ref{fig:fig1}(a). The data show a single
resolution limited mode propagating along the $(hh\frac{3}{2})$ direction between energy transfers $\hbar\omega$
of 1 and 3 meV.  The bandwidth and gap of the spectrum is in good agreement with values
determined from powder INS measurements \cite{stonebamno}.  Measurements at several high-symmetry
locations in the $(hk0)$ scattering plane were also performed.  These were made in the identical
instrumental configuration with the focusing analyzer broadened to $\approx 23.1$~cm to integrate
over additional scattering intensity.  The broadened lineshape for $(\frac{1}{2}00)$
in Fig.~\ref{fig:fig1}(a) reflects the broadened $Q$ resolution of the wider analyzer.  As shown in Fig.~\ref{fig:fig1}(a), the measured scattering intensity in the $(hk0)$ plane is very weak compared
to the $(hhl)$ plane.

To illustrate the dispersion and wave vector dependent scattering intensity, we combine a series of constant
$\mathbf{Q}$ scans in the image shown in Fig.~\ref{fig:fig2}(a)-(d).  Figure~\ref{fig:fig2}(i) shows the
path through wave vector space for these data.  Figure~\ref{fig:fig2} indicates a single dispersive mode in
both the $(hh\zeta)$ and $(\zeta\zeta l)$ directions.  The dispersion along the $(hh\zeta)$ wave vectors has
a distinct \textsf{W} shape with clear minima in the vicinity of $h\approx \frac{1}{3}$ and $h\approx \frac{2}{3}$
and a local maximum at the $h=\frac{1}{2}$.  The dispersion along the $l$ wave vector has a periodicity of three reciprocal lattice units (rlu).

\begin{figure}[tb!]
\centering\includegraphics[scale=0.765]{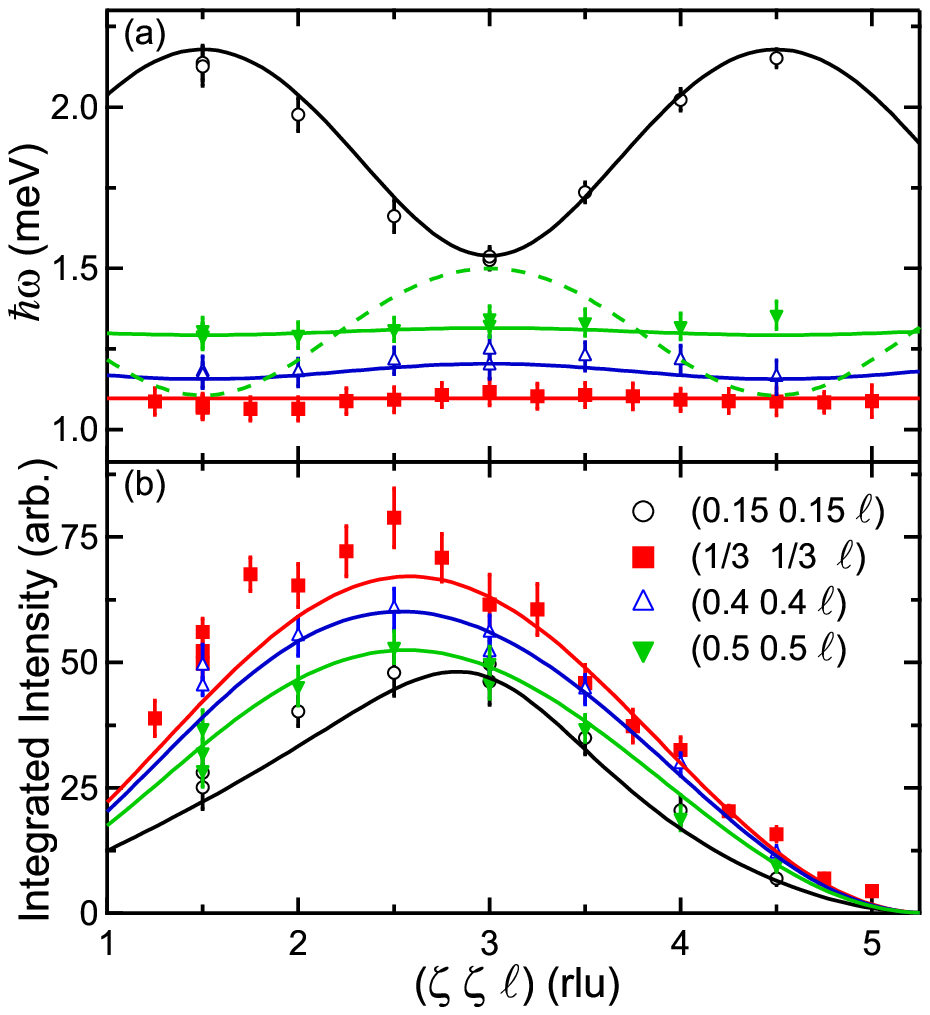}
\caption{\label{fig:fig3}(a) Dispersion and (b) integrated scattering intensity of Ba$_3$Mn$_2$O$_8$ determined
from Gaussian fits to constant $\mathbf{Q}$ scans.  Solid lines in (a) are fits to the dispersion as discussed
in the text.  Dashed line in (a)
is the fitted dispersion for $(\frac{1}{2}\frac{1}{2}l)$ fixing $J_4=0$ resulting in exchange constants $J_0=1.640(8)$,~$J_1=-0.117(5)$~and $J_2-J_3=0.109(2)$.  Solid lines in (b) are fits to the scattering intensity as described in the text.}
\end{figure}

126 individual constant $\mathbf{Q}$ scans were separately fit to a single inelastic Gaussian peak.
The peak positions along several high-symmetry directions are plotted in Fig.~\ref{fig:fig2}(a)-(d) and along several $(\zeta\zeta l)$
wave vectors in Fig.~\ref{fig:fig3}(a).  The singlet-triplet excitation is fairly dispersive along the $(00l)$ direction as shown in
Fig.~\ref{fig:fig2}(d) and also along the $l$ direction near the zone boundary as shown in Fig.~\ref{fig:fig3}(a) for $(0.15~0.15~l)$.
However, there is almost no dispersion along $l$ for $0.3\leq h\leq0.7$.  In fact for the $(\zeta \zeta l)$ scans shown in Fig.~\ref{fig:fig3}(a),
the dispersion flattens and continuously changes phase as one moves from $\zeta=0$ to $\zeta=\frac{1}{2}$.

The three original exchange constants determined from analysis of powder INS data do not account for the
lack of dispersion along the $l$ direction near the zone center.  Rather we must include the NNN inter-dimer
interaction in the Heisenberg Hamiltonian $J_4$.  As in the powder analysis, we use a dispersion based upon weakly coupled $S=1$ dimers:
\begin{equation}
\label{eq:rpadisp}
\hbar\omega^{\nu}(\mathbf{Q}) = \sqrt{\Delta_{\nu}^{2} + \frac{8}{3}\Delta_{\nu}\mathcal{J}(\mathbf{Q})R(T)}
\end{equation}
where $\nu=\{1,0,-1\}$ denotes the spin projection $S^z$ of the triplet excitation, $\mathcal{J}(\mathbf{Q})$ is the Fourier sum of interactions beyond
dimer exchange, $\Delta_{0} = J_0+2D/3$, $\Delta_{\pm 1}  = J_0 -D/3$ and $R(T)$ is
the thermal population difference between ground and excited states.  For $S=1$ antiferromagnetic dimers,
\begin{equation}
\label{eq:roftdisp}
R(T) = \frac{1-\exp(-J_0 \beta)}{1+3\exp(-J_0 \beta) + 5\exp(-3 J_0 \beta)},
\end{equation}
where $\beta = \frac{1}{k_B T}$.
Including the $J_4$ interaction, the Fourier sum becomes
\begin{eqnarray}
\label{eq:dispeqn}
\nonumber \omega_1 &=& \cos(\frac{2\pi}{3}[-h+k+l]) + \cos(\frac{2\pi}{3}[-h-2k+l]) \\
\nonumber & &
+ \cos(\frac{2\pi}{3}[2h + k +l])  \\
\nonumber \omega_2 &=& \cos(2\pi k) + \cos(2\pi[h+k]) + \cos(2\pi h) \\
\nonumber \omega_4 &=& \cos(\frac{2\pi}{3}[2h - 2k + l]) + \cos(\frac{2\pi}{3}[2h + 4k + l]) \\
\nonumber & & + \cos(\frac{2\pi}{3}[-4h -2k +l])\\
\mathcal{J}(\mathbf{Q}) &=& J_1\omega_{1} + 2(J_2-J_3)\omega_2 + J_4\omega_{4}.
\end{eqnarray}
 The small anisotropy $D=-0.032$ meV \cite{samulonbamno,stonebamno}has been determined from zero-field ESR experiments \cite{esrreff}. For Ba$_3$Mn$_2$O$_8$ in zero magnetic field, $D$ will not change the dispersion and will only
broaden the constant $\mathbf{Q}$ scans.
As shown in Fig.~\ref{fig:fig1}(a), $J_4$ corresponds to NNN interdimer interactions along the $c$-axis between bilayers.
This exchange path is through the 12-fold coordinated Ba sites which are co-planer with the O sites associated with the Mn tetrahedra.
Each of the 12-fold coordinated Ba sites participates in three such $J_4$ exchange paths.  A simultaneous fit of the peak positions of
the constant $\mathbf{Q}$ scans in the $(hhl)$ scattering plane yields $J_0 = 1.642(3)$, $J_1 = -0.118(2)$, $(J_2 - J_3) = 0.1136(7)$
and $J_4 = -0.037(2)$~meV.  $J_2$ and $J_3$ represent interdimer exchange between the same spin dimers.   They cannot be independently
determined from the zero-field dispersion because $\mathcal{J}(\mathbf{Q})$ depends on their difference. The phase shift for triplets
propagating between dimers via $J_2$ or $J_3$ (see Fig.~\ref{fig:fig1}(b)) is due to the opposite symmetries of the singlet (symmetric) and the
triplet (antisymmetric) states under a permutation of the two sites that form the dimer.
The corresponding dispersion curves shown
in Figs.~\ref{fig:fig2}(a)-(d) and ~\ref{fig:fig3}(a) agree very well with the measurement.  We also plot
the fitted dispersion along $(\frac{1}{2}\frac{1}{2}l)$ for $J_4=0$ (dashed line in Fig.~\ref{fig:fig3}(a).  It is clear that the NNN exchange is needed to describe
the measured dispersion.  The dispersion minimum  $\hbar\omega = 1.081$~meV occurs at wave vectors $(\frac{1}{3}-\delta, \frac{1}{3}-\delta, l)$
and $(\frac{2}{3}+\delta, \frac{2}{3}+\delta, l)$ with $\delta \approx 0.0259$.  This value is similar to that determined from our powder measurements,
and recent mean-field calculations\cite{xuprb2008}.

Figures~\ref{fig:fig2}(e)-(h) and~\ref{fig:fig3}(b) show the $\mathbf{Q}$ dependent integrated scattering intensity.
Based upon the established SMA dimer model\cite{stonebamno}, the magnetic scattering intensity for $k_B T\ll J_0$is
\begin{equation}
\label{eq:fullsofqwsma}
I_m(\mathbf{Q},\omega) \propto |F(Q)|^2 \frac{1-\cos({\mathbf{Q}}\cdot \mathbf{d})}
{\hbar\omega(\mathbf{Q})}\delta(\hbar\omega - \hbar\omega(\mathbf{Q})),
\end{equation}
where $|F(Q)|$ is the magnetic form factor \cite{magformfacnote,pjbrown} and $\mathbf{d}$ is the intradimer vector:
$\mathbf{d} = 0.1858 \mathbf{c}$.  The dimer structure factor explains the lack of intensity found in the $(hk0)$
plane, \textit{i.e.}~$\mathbf{Q}_{(hk0)}\perp\mathbf{d}$.  We fit the integrated intensity to the SMA model in
Eq.~(\ref{eq:fullsofqwsma}) using the dispersion found earlier and only a single multiplicative fitting parameter.
The fit agrees quite well with the measured results.  The dimer structure factor accounts for almost all of the scattering intensity.

The interdimer exchange corresponding to triplet propagation out
of the hexagonal $ab$-plane is ferromagnetic for both $J_1$ and $J_4$,
whereas interdimer coupling in the hexagonal $ab$-plane is entirely antiferromagnetic and consequently geometrically frustrated. We can gain insight into the consequences of the two interlayer couplings by considering the magnetic structure for $H_{c1}<H<H_{c2}$.  The in-plane geometric frustration alone leads to a $120^{\circ}$ magnetic structure for a single isolated bilayer.  Such a structure blocks the inter-layer couplings $J_1$ and $J_4$ because a spin ${\bf S}_i$ on a
given bilayer interacts with the sum of three spins that are on the three different sublattices (A, B and C) of the adjacent bilayer
(this sum is zero for a $120^{\circ}$ structure). Consequently, the ferromagnetic interaction $J_1$ induces a spiral phase
($\alpha \neq 120^{\circ}$) \cite{uchidaprb2002,samulonbamno} to have a net moment on each triangle.
This moment is ferromagneneticaly coupled to the spin ${\bf S}_i$ that is at the center of the corresponding triangle
(and in the next bilayer). However, such spiral distortion is not favorable for the ferromagnetic $J_4$ term: the three spins
that are next-nearest-neighbors of ${\bf S}_i$ in the next bilayer form a bigger triangle and the net moment of this triangle
is {\it antiferromagnetically} aligned with ${\bf S}_i$. Therefore, the $J_1$ exchange interaction competes against $J_4$ producing a
flatter singlet-triplet dispersion along the $c$-axis, i.e., an effective quasi-2D like behavior.
This competition also leads to smaller deviation of $\alpha$ from $120^{\circ}$.

Based upon the determined exchange constants and an appropriate choice of $J_{3}=0.167$~meV such that $H_{c1}$ agrees with its well determined experimental value, we are able to determine the remaining critical fields.
Using a two-level approach \cite{uchidaprb2002}
keeping only the singlet ($S^z=2$ quintet) and the $S^z=1$ triplet dimer states for
$H_{c1} < H < H_{c2}$ ($H_{c3} < H < H_{c4}$), we obtain
values for the four critical fields:
$H_{c1}=10$~T, $H_{c2}=27$~T; $H_{c3}=32.75$~T; $H_{c4}=47.75$~T.
The agreement with the experimental values is quite good
considering that neglecting the excited triplet $S^z=0,-1$ states introduces errors of the order of $1$~T.
For instance, if we include the effect of quadratic fluctuations to the $S^z=0,-1$ triplet states,
the lowest critical field becomes$H_{c1}=\omega^{1}({\bf Q}_{min}) / g \mu_B = 9$~T in excellent
agreement with the experimental value [${\bf Q}_{min}$ is the wave-vector that minimizes $\omega^{1}({\bf Q})$].

Through single crystal INS measurements of Ba$_3$Mn$_2$O$_8$ we find that NNN exchange interactions are
significant in describing the system.  The nearest and NNN inter-bilayer couplings are both ferromagnetic.
It is clear that the antiferromagnetic interactions within the bilayer lead to a geometrically frustrated system.  However for Ba$_3$Mn$_2$O$_8$ we find a novel case where ferromagnetic inter-bilayer exchange actually serves to enhance the quasi-2D behavior of the bilayer.   The refined exchange constants also provide a zero-field confirmation of the high magnetic field QCPs.  In addition to marked progress in understanding the quantum critical phase diagram of Ba$_3$Mn$_2$O$_8$ and its excitations and ordered states, the determined dispersion should also be applicable to other isostructural gapped magnetic compounds in the $A_3Cr_2$O$_8$ series where $A=$Ca$^{2+}$, Sr$^{2+}$ or Ba$^{2+}$  such as Ba$_3$Cr$_2$O$_8$ \cite{isostruct4, isostruct1, isostruct6, isostruct7} and Sr$_3$Cr$_2$O$_8$ \cite{isostruct5, isostruct9}.

The research at Oak Ridge National Laboratory was sponsored by the Scientific User Facilities Division, Office
of Basic Energy Sciences, U. S. Department of Energy.  This work utilized facilities supported in part by the National
Science Foundation under Agreement No. DMR-0454672.  work at Stanford was supported by the National Science Foundation,
under grant DMR 0705087.

\end{document}